# An Alternative Scheme for Calculating the Unrestricted Hartree-Fock Equation: Application to the Boron and Neon Atoms

Mitiyasu MIYASITA[1,2], Katsuhiko HIGUCHI[1], Masahiko HIGUCHI[3]

[1]*Graduate School of Advanced Science of Matter, Hiroshima University, Higashi-Hiroshima 739-8527, Japan*

[2]*CCSE, Japan Atomic Energy Agency, 6-9-3 Higashi-Ueno, Taito-ku, Tokyo 110-0015, Japan*

[3]*Department of Physics, Faculty of Science, Shinshu University, Matsumoto 390-8621, Japan*

(June 15, 2009)

**Abstract**

We present an alternative scheme for calculating the unrestricted Hartree-Fock equation. The scheme is based on the variational method utilizing the sophisticated basis functions that include no adjustable parameters.  The validity and accuracy of the present scheme are confirmed by actual calculations of the boron and neon atoms.  It is shown that the present scheme not only gives the reasonably lower total energy but also conserves the virial relation with enough accuracy.

## §1  Introduction

The single-particle wave function and spectra for the atomic system are fundamental for understanding the electronic and magnetic properties of solids.   For instance, hopping and Coulomb integrals included in the model Hamiltonian[1,2] and LDA+$U$ method[3,4], etc, are defined using the atomic wave functions.   Also in the tight-binding method,[5] the matrix elements of the Hamiltonian are constructed from the atomic wave functions and spectra.   The electronic and magnetic properties that are related to the well-localized electrons of solids are necessarily related to the atomic wave functions and spectra.   Thus, there is no doubt that appropriate description for the atomic system will lead to better understanding of solids.

Concerning the theoretical framework for calculating the atomic structures, the most standard one is the Hartree-Fock (HF) equation.[6,7]   The HF equation does not contain the correlation effects at all, but such correlation effects are small in the atomic system compared to the solids.   The ratio of the correlation energy to the exchange energy is roughly 1/10 as large as that of solids.[8,9,10]   Therefore, the HF equation is a good starting point for describing the atomic structures.

There exist two kinds of HF equations, i.e., restricted Hartree-Fock (RHF) equation and unrestricted Hartree-Fock (UHF) equation.[11]   The former has the merit of conserving the total spin moment rigorously, but has the demerit concerning the matrix type of the Lagrange multipliers. The Lagrange multipliers generally appear in the single-particle equation in a form of the matrix elements by imposing the orthonormality on the orbitals.   Such the matrix cannot always be transformed into the diagonal form in the case of the RHF equation.[12]   This means that the RHF equation cannot always be transformed into the canonical form.   These nondiagonal elements of the Lagrange multipliers are small but can not be neglected in some case.[12,13]   If the nondiagonal elements of the Lagrange multipliers are neglected for the practical aim, one can utilize the so-called term-dependent HF equation, which is derived by taking the functional derivatives of the total energy of a single term with respect to the radial functions.[12,14]   It is shown by Slater that this scheme is equivalent to the approximation of taking the spherical average of the potentials.[14,15] Extended version of this approximation has been also developed by Slater.[13]   It starts with the total energy that is averaged with respect to all the allowed terms to some electron configuration. The resultant RHF equation, which may be called the term-averaged HF equation, has been numerically solved by Fischer et al.[16]   The results have been regarded as a standard reference data of the RHF equation.   In addition to the above methods, there is an effective method to solve the RHF equation, i.e., the Hartree-Fock-Roothaan (HF-R) method.[17,18,19]   In the HF-R method, the solutions are expanded with the empirical basis functions such as the Gaussian-type orbital (GTO) and Slater-type orbital (STO).   The problem of solving the RHF equation results in the generalized eigenvalue problem, which is a strong advantage from the viewpoint of the numerical calculations. The HF-R method usually deals with the canonical RHF equation. In this sense, the HF-R method also gives an approximation of the RHF equation.

Next let us sketch the features of the latter one (UHF equation). The UHF method has the merit that the equation can take the canonical form rigorously, but it has the demerits that the total spin moment is not always guaranteed to be conserved. Due to such the demerit, the UHF method has not been actively treated, as compared to the RHF method.[11,20-24] However, the UHF equation is obviously more natural and reasonable than the RHF equation because the spatial wave functions for the up-spin and down-spin states are generally different from each other. If we deal with the UHF equation appropriately, the total energy would be lower than that of the RHF method.

In this paper, we present an alternative scheme for calculating the UHF equation, which is based on the variational method. As the expansion basis, we choose the solutions of the Xα method with the spherically-averaged potentials[25]. This choice of the basis functions reduces the computational task in terms of the number of the expansion basis compared to the HF-R method. Furthermore, as shown in the subsequent sections, the present scheme not only improves the total energy but also conserves the virial relation with enough accuracy. Of course, the present scheme includes the effects of the nonspherical distribution of electrons. This effect can not be negligible for the atomic structures, as already indicated by Slater[26] and ours[27,28], and unambiguously shown by our preceding works.[29,30]

Organization of this paper is as follows. In Sec. 2, we explain the outline of the present scheme. The results of the test calculation for the boron and neon atoms, and corresponding discussions are given in Sec. 3. Finally, in Sec. 4, we summarize the results and give some comments on it.

## §2  Variational Method

In this section, we present a variational method for calculating the UHF equation by means of sophisticated basis functions. Let us start with the UHF single-particle equation:

$$-\frac{\hbar^2}{2m}\nabla^2\psi_{i\sigma_i}(\mathbf{r}) - \frac{Ze^2}{r}\psi_{i\sigma_i}(\mathbf{r})$$
$$+e^2\sum_{j\sigma_j}^{\text{occ.}}\int\frac{|\psi_{j\sigma_j}(\mathbf{r}')|^2}{|\mathbf{r}-\mathbf{r}'|}d\mathbf{r}'\psi_{i\sigma_i}(\mathbf{r}) - e^2\delta_{\sigma_i\sigma_j}\sum_{j\sigma_j}^{\text{occ.}}\int\frac{\psi_{j\sigma_j}^*(\mathbf{r}')\psi_{i\sigma_i}(\mathbf{r}')}{|\mathbf{r}-\mathbf{r}'|}d\mathbf{r}'\psi_{j\sigma_j}(\mathbf{r})$$
$$=\varepsilon_{i\sigma_i}\psi_{i\sigma_i}(\mathbf{r}),$$

(1)

where $\sigma$ denotes the up-spin or down-spin. The basic idea of our scheme is almost the same as that of our recent work[29]. Namely, the solution of eq. (1) is expanding with the set of known functions as basis functions. It should be noted that the variational method always includes the arbitrariness of choosing basis functions and the choice affects calculation results directly.

As the basis functions, we adopt the eigenfunction for the Hamiltonian defined by

$$\hat{H}_0^\sigma := -\frac{\hbar^2}{2m}\nabla^2 - \frac{Ze^2}{r} + V_S^H(r) + V_{S,\sigma}^{X\alpha,ex}(r), \tag{2}$$

where $V_S^H(r)$ and $V_{S,\sigma}^{X\alpha,ex}(r)$ stand for the spherical part of Hartree and exchange potential of the Xα method[25], respectively. Utilizing the numerical methods such as the Herman-Skillman method[31], the eigenfunctions and eigenvalues of the eq. (2) can be easily obtained, which are denoted as $p_{nl}^\sigma(r)Y_{lm}(\theta,\varphi)$ and $\varepsilon_{nl}^\sigma$, respectively. We shall expand the solution of eq. (1) with respect to such the eigenfunctions:

$$\psi_{i\sigma_i}(\mathbf{r}) = \sum_{nlm} C_{nlm}^{i\sigma_i} p_{nl}^{\sigma_i}(r) Y_{lm}(\theta,\varphi). \tag{3}$$

Substituting eqs. (2) and (3) into eq. (1), and writing distinctly the spherical part and the rest one (which is denoted by $V_{NS}^H(\mathbf{r})$) of the Hartree potential, we get

$$\left\{\hat{H}_0^{\sigma_i} + V_{NS}^H(\mathbf{r}) - V_{S,\sigma_i}^{X\alpha,ex}(r)\right\} \sum_{nlm} C_{nlm}^{i\sigma_i} p_{nl}^{\sigma_i}(r) Y_{lm}(\theta,\varphi)$$
$$- e^2 \delta_{\sigma_i\sigma_j} \sum_{j\sigma_j}^{occ.} \int \frac{\psi_{j\sigma_j}^*(\mathbf{r}')\sum_{nlm} C_{nlm}^{i\sigma_i} p_{nl}^{\sigma_i}(r')Y_{lm}(\theta',\varphi')}{|\mathbf{r}-\mathbf{r}'|} d\mathbf{r}' \psi_{j\sigma_j}(\mathbf{r}) = \varepsilon_{i\sigma_i} \sum_{nlm} C_{nlm}^{i\sigma_i} p_{nl}^{\sigma_i}(r)Y_{lm}(\theta,\varphi).$$

$$\tag{4}$$

Multiplying $p_{n_a l_a}^{\sigma_a *}(r) Y_{l_a m_a}^*(\theta,\varphi)$ on both sides of eq. (4) and integrating over the whole space, we have

$$\sum_{nlm} C_{nlm}^{i\sigma_i} \left\{ \delta_{l_a l} \delta_{m_a m} O_{n_a l_a \sigma_a nl\sigma_i} \left( \frac{\varepsilon_{n_a l_a}^{\sigma_a 0} + \varepsilon_{nl}^{\sigma_i 0}}{2} - \varepsilon_{i\sigma_i} \right) + V_{n_a l_a m_a \sigma_a nlm\sigma_i} - \sum_{j\sigma_j}^{\text{occ.}} \delta_{\sigma_i \sigma_j} E_{n_a l_a m_a \sigma_a nlm\sigma_i}^{j\sigma_j} \right\} = 0$$

(5)

where

$$O_{n_a l_a \sigma_a nl\sigma_i} = \int p_{n_a l_a}^{\sigma_a *}(r) p_{nl}^{\sigma_i}(r) r^2 dr, \tag{6}$$

$$V_{n_a l_a m_a \sigma_a nlm\sigma_i} = \int p_{n_a l_a}^{\sigma_a *}(r) Y_{l_a m_a}^*(\theta, \varphi) \{V_{\text{NS}}^{\text{H}}(\mathbf{r}) - V_{\text{S},\sigma_i}^{\text{X}\alpha,ex}(r)\} p_{nl}^{\sigma_i}(r) Y_{lm}(\theta, \varphi) d\mathbf{r}, \tag{7}$$

$$E_{n_a l_a m_a \sigma_a nlm\sigma_i}^{j\sigma_j} = e^2 \int p_{n_a l_a}^{\sigma_a *}(r) Y_{l_a m_a}^*(\theta, \varphi) \frac{\phi_{j\sigma_j}^*(\mathbf{r}') p_{nl}^{\sigma_i}(r') Y_{lm}(\theta', \varphi')}{|\mathbf{r} - \mathbf{r}'|} \phi_{j\sigma_j}(\mathbf{r}) d\mathbf{r} d\mathbf{r}', \tag{8}$$

and $\phi_{i\sigma_i}(\mathbf{r})$ is the result of the previous iteration. Equation (5) is just the generalized eigenvalue problem. If the matrix elements, $O_{n_a l_a \sigma_a nl\sigma_i}$, $V_{n_a l_a m_a \sigma_a nlm\sigma_i}$, $E_{n_a l_a m_a \sigma_a nlm\sigma_i}^{j\sigma_j}$ and the energy spectra of the spherical approximation, $\varepsilon_{i\sigma_i}^0$, are given, then we can obtain the eigenvalues, $\varepsilon_{i\sigma_i}$, and eigenfunctions, $\{C_{nlm}^{i\sigma_i}\}$. It should be noted that the eigenvalues $\varepsilon_{i\sigma_i}$ are guaranteed to be real since two matrices included in eq. (5) are hermitian. The angular integration in eqs. (7) and (8) can be analytically calculated by using the Wigner 3j-symbols.[32]

The potentials eq. (7) should be determined in a self-consistent way[6]. The corresponding basis functions in eq. (3) are modified for each iteration since the function $p_{nl}^{\sigma}(r)$ is the radial part of solution for the Hamiltonian (2). This is the striking feature of the present scheme, which enables us to solve the UHF equation more rapidly with not so many of basis functions. The iteration is continued until self-consistency for the potentials is achieved.

## §3 Results and Discussions

In order confirm the effectiveness of our scheme, actual calculations are applied to the boron and neon atoms. The number of basis functions of eq. (3) is 23, which have the following quantum

numbers:

$$\begin{aligned}(nlm) = &(100), (200), (211), (210), (21-1),\\ &(300), (311), (310), (31-1), (322), (321), (320), (32-1), (32-2),\\ &(400), (411), (410), (41-1), (422), (421), (420), (42-1), (42-2).\end{aligned}$$

As mentioned above, we use the spherical approximated Xα method for preparing the basis functions. The parameter α is determined by requiring that the virial theorem holds. The most optimized values are 0.8442691 for boron and 0.776 for neon, respectively. (See Figure 1)

Let us give the results of the test calculation for the boron and neon atoms. Figure 2 shows the energy spectra of the present scheme, together with those of conventional ones[16,19]. The differences between ours and conventional ones are not small, especially for the outer states.

Table 1 shows the total energy of the present and conventional schemes. Present scheme gives the lower total energy than that of the conventional ones. This is not only because the UHF has no restriction on the wavefunction such that the spatial parts are identical for the up- and down-spin states, but also-because the choice of the basis functions in more appropriate than the conventional HF scheme. The latter reason is confirmed by the fact that the total energy of the neon atom that has the same spatial wavefunction for the up- and down-spin states due to no spin-polarization becomes lower than the conventional RHF scheme. Furthermore, we would like to stress that the virial relation is conserved with enough accuracy in the present scheme.

## §4 Concluding Remarks

In this paper, we present an alternative scheme for calculating the UHF equation. The validity and accuracy of this scheme have been confirmed by actual calculations for the boron and neon atoms. The present scheme can lower the total energy reasonably, while conserving the virial relation.

The present scheme clearly shows the necessity of modifying the single-particle picture of atomic systems. Furthermore, we can say that the present scheme may open the possibility of improving the estimation of the various parameters concerning the solid state physics, which are based on the atomic single-particle wave functions and spectra.

For the future work, we have to consider the relativistic and/or correlation effects in order to improve the single-particle picture beyond the HF approximation. These are completely neglected in the present calculations. Especially concerning the relativistic effect, it seems to be indispensable for describing the electric structures of the heavier atoms.

**Acknowledgements**

This work was partially supported by Grant-in-Aid for Scientific Research (No. 19540399) and for Scientific Research in Priority Areas (No. 17064006) of The Ministry of Education, Culture, Sports, Science, and Technology, Japan.

**Figure captions**

Fig. 1.   Parameter alpha and virial ratio for boron and neon atoms.   Open and black circle denotes born and neon atoms, respectively.

Fig. 2.   Energy spectra for the boron and neon atoms.   The first and third columns show the results for the conventional HF method which is calculated by E. Clementi and C. Roetti or C. F. Fischer[16,19].   The second and forth columns are the results for the present scheme.   The up- and down-arrows denote the occupied states, and open circle the unoccupied states.   All values are given in Rydberg Unit.

**Tables 1**

**Table 1.** The total energies of the conventional result[16,19] and the present one. All values are given in Rydberg Unit.

| Atom  | Conventional result | Present result |
|-------|---------------------|----------------|
| boron | -49.058             | -49.194        |
| neon  | -257.094            | -257.335       |

**Figure 1**

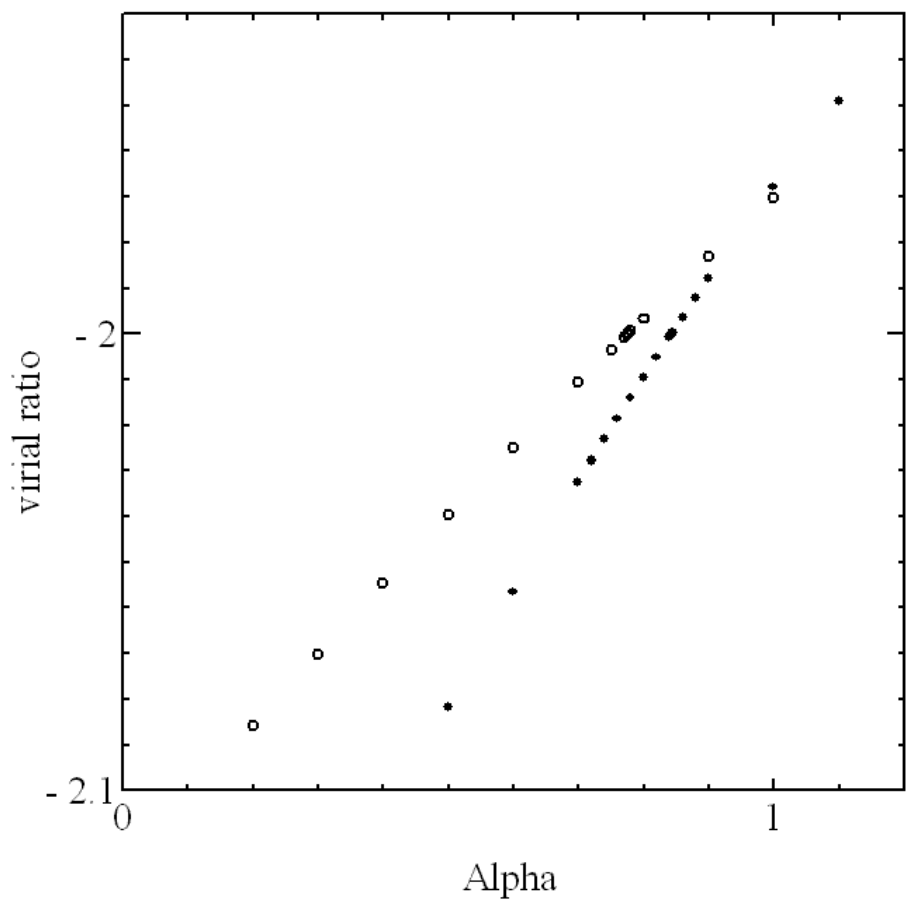

**Figure 2**

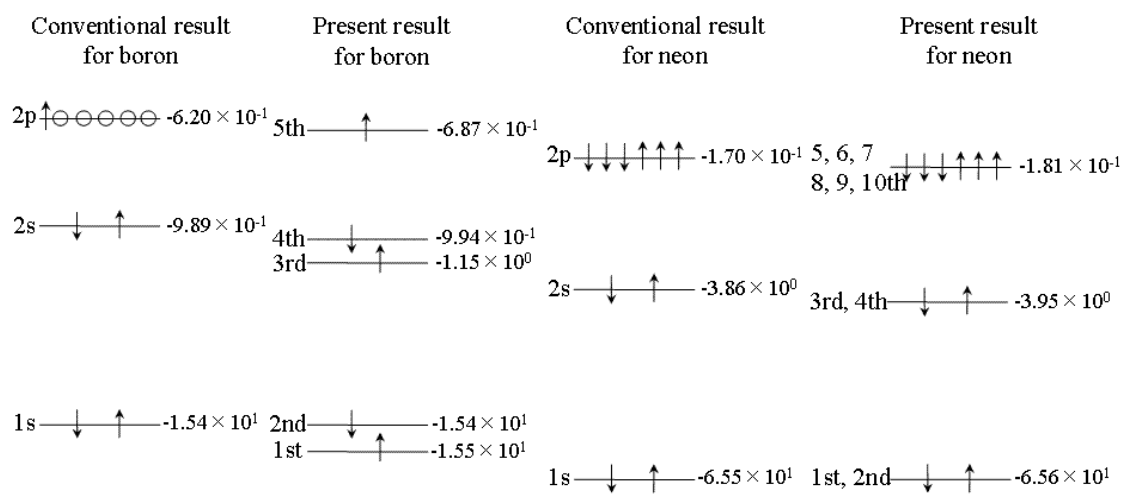